\documentclass[prl,showpacs,superscriptaddress,amsfonts,amsmath,floatfix,twocolumn]{revtex4-1}
\usepackage{graphicx}
\usepackage[dvipdfm]{hyperref} 
\usepackage{verbatim} 
\usepackage[normalem]{ulem}
\usepackage{color}

\newcommand{\ket}[1]{|#1\rangle}

\newcommand{\add}[1]{\textcolor{black}{#1}}

\newcommand{\affA}{Center for Integrated Quantum Science and Technology, Institute for Complex Quantum Systems, Universit\"at Ulm, D-89069 Ulm, Germany}
\newcommand{\affB}{Theoretische Chemie, Physikalisch-Chemisches Institut, Universit\"at Heidelberg, D-69120 Heidelberg, Germany}
\newcommand{\affC}{Zentrum f\"ur Optische Quantentechnologien and The Hamburg Centre for Ultrafast Imaging, Universit\"at Hamburg, Luruper Chaussee 149, D-22761 Hamburg, Germany}
\newcommand{\affD}{ICFO-Institut de Ciencies Fotoniques, Mediterranean Technology Park, S-08860 Castelldefels (Barcelona), Spain}

\begin{document}
\title{Quantum Speed Limit and Optimal Control of Many-Boson Dynamics}
\author{Ioannis Brouzos}\affiliation{\affA}
\author{Alexej I. Streltsov}\affiliation{\affB}
\author{Antonio Negretti}\affiliation{\affC}
\author{Ressa S. Said}\affiliation{\affA}
\author{Tommaso Caneva}\affiliation{\affD}
\author{Simone Montangero}\affiliation{\affA}
\author{Tommaso Calarco}\affiliation{\affA}

\date{\today}
\begin{abstract}
We extend the concept of quantum speed limit --  the minimal time needed to perform a driven evolution
-- to complex interacting many-body systems.
We investigate a prototypical many-body  system\add{,} a bosonic Josephson junction\add{,} at increasing levels of complexity:
(a) within the two-mode approximation {corresponding to} a nonlinear two-level system,
(b) at the mean-field level \add{by} solving the nonlinear Gross-Pitaevskii equation in a double well potential, and 
(c) at an exact many-body level \add{by} solving the time-dependent many-body Schr\"odinger equation.
We propose a control protocol to transfer atoms from {the ground state of a well to the ground state of the neighbouring well.}
{Furthermore, we} show that \add{the} detrimental effects of the \add{inter-particle} repulsion can be {eliminated} by 
\add{means of a} compensating control pulse, yielding\add{, quite surprisingly,} an enhancement of the transfer speed 
{because of the particle interaction} -- in contrast to the self-trapping scenario.
{Finally, we perform numerical optimisations of} both \add{the} nonlinear and 
\add{the (exact)} many-body \add{quantum} dynamics {in order to further enhance the transfer efficiency} close to the quantum speed limit.
\end{abstract}

\maketitle

The maximum speed of evolution of a quantum system has been a subject of \add{theoretical} investigation since decades. This research activity 
-- mainly focused on state-to-state transfer for single-particle quantum dynamics -- resulted in the introduction of the \add{concept generally known as}
quantum speed limit (QSL), \add{namely} the minimum time needed to achieve a certain quantum state transformation~\cite{qsl_lit1,qsl_lit2}. 
In the last \add{few} years, this fascinating theoretical concept\add{, which} {arises} from an energy-time uncertainty relation, has started to have practical relevance 
{especially for the development of} quantum technologies {and because quantum 
gas laboratories are now capable {of}} probing it~\cite{qsl_exp1,qsl_exp2}. 
Indeed, recent experiments with cold atoms on a chip reported the achievement of time-optimal processes at the QSL~\cite{qsl_exp2}. 
However, the QSL for nonlinear and many-body quantum systems has not been clearly defined yet and, in particular,  
the influence of the inter-particle interaction 
on the QSL has to be still clarified.
Here, we provide an exhaustive theoretical analysis on this subject and we apply it to the relevant scenario of ultracold bosons in a double well 
that has been realized~\cite{bjj_exp1,bjj_exp2} and controlled~\cite{bjj_exp3} experimentally. 
We characterize the QSL of nonlinear and many-body systems of bosons,  and show that optimal control might be effectively applied to 
achieve time-optimal transformations. 
In particular, we show that optimal control strategies can be developed to cancel the decelerating effects of the repulsive atom interactions. 
These strategies differ substantially in the vicinity of the QSL in such a way that the combination of an 
exact many-body quantum description and optimal control is instrumental in order to achieve the best performance.
 Our optimal control schemes are characterized by simplicity of the pulses and feasibility of their application in all experiments dealing with linear or nonlinear two-level systems. They contribute to solving
key problems of the atom transfer by compensating self-trapping and damping of the oscillations in a time-optimal manner, and achieve complete control of the motion on the Bloch sphere which facilitates interferometric protocols.

\begin{figure}[tp]
\setlength{\unitlength}{.9cm}
\begin{picture}(8.5,12.5)
\put(-0.6,-1.9){\includegraphics[scale=0.4]{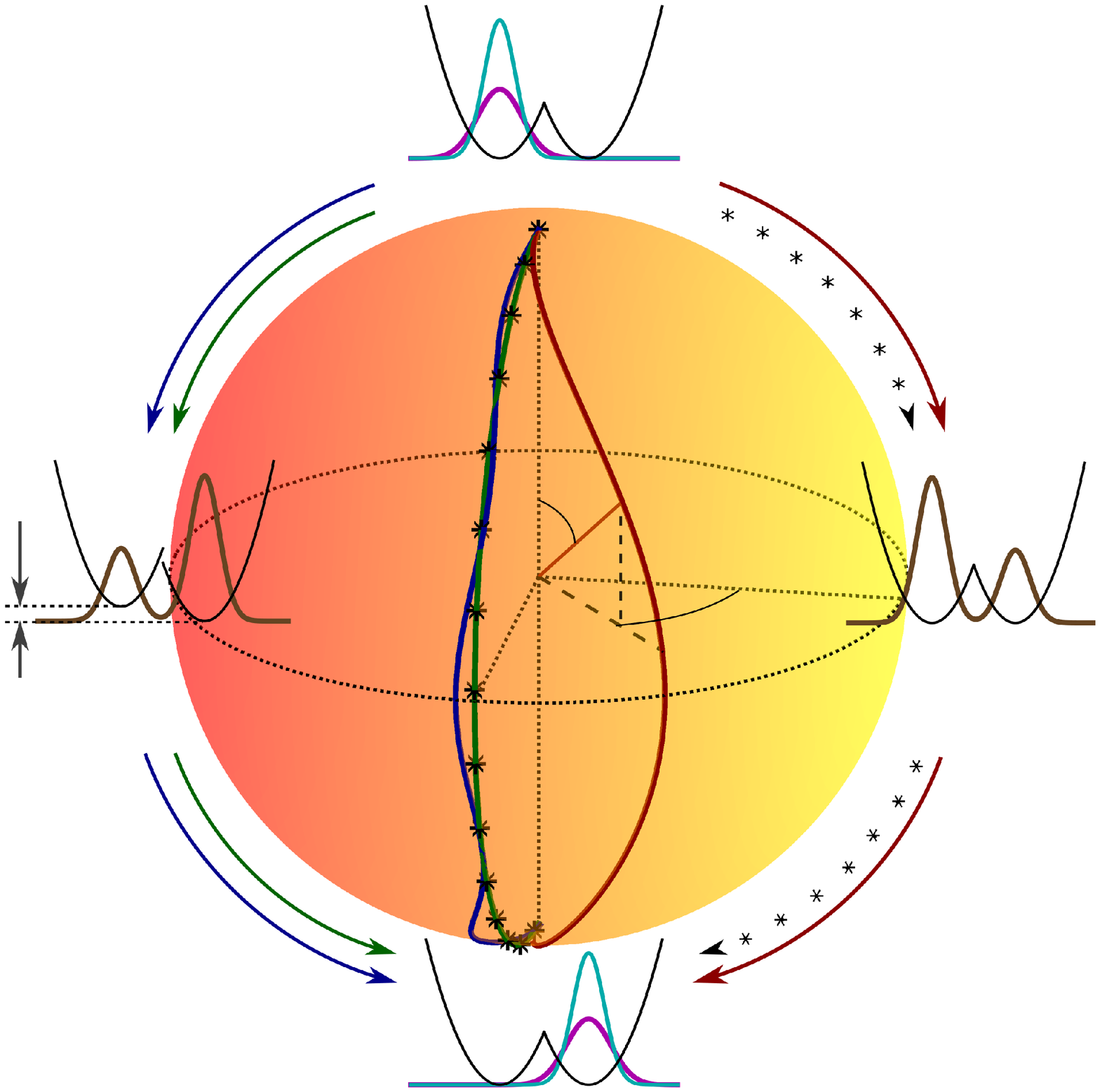}}
\put(-0.1,8.3){{(d)}}
\put(3.8,7.4){\bf{Initial}}
\put(-0.2,3.3){\bf{Controlled}}
\put(6.3,3.5){\bf{Uncontrolled}}
\put(3.8,0.1){\bf{Target}}
\put(5.5,3.9){$\phi$}
\put(4.6,5.0){$\theta$}
\put(-0.2,9.6){\includegraphics[scale=0.30]{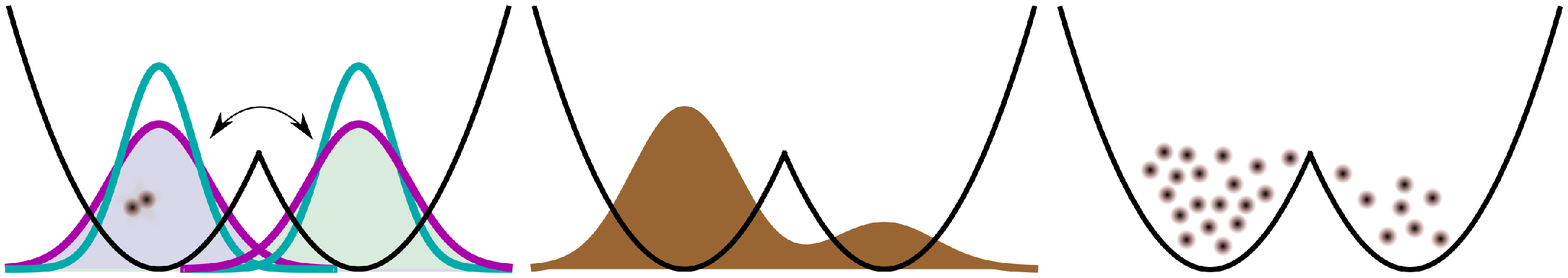}}
\put(-0.1,11.8){(a) \bf{Two-mode}}
\put(0.5,11.2){CCP Theory}
\put(3.0,11.8){(b) \bf{Mean-field}}
\put(3.1,11.2){Nonlinear GPE}
\put(6.1,11.8){(c) \bf{Many-body}}
\put(6.4,11.2){MCTDHB}
\put(0.6,9.8){ $U$}
\put(1.2,10.7){$J$}
\end{picture}
\caption{(a)-(c) The \add{dynamics of the bosonic Josephson junction} is studied at various levels of complexity and approximations.
 (d) Bloch Sphere representation of different paths for the two-mode nonlinear dynamics for $UN/2J=1$:
uncontrolled GPE (red), CCP (green),
and CRAB-optimized with constraints on the pulse $D(0)=-D(T)=2J$ (blue). \add{The path with black asterisks} represents a geodesic.
Upper and lower sketches: Initial and target states  in the double-well potential (black line) for strong (magenta) and weak (cyan) interaction strength.
Left and right sketches: controlled and uncontrolled wave functions and \add{trap} potential\add{s}.\vspace{-0.6cm}
}
\label{fig1}
\end{figure}

\add{In our study} different degrees of complexity in the description of {the} system are gradually {considered} (see Fig.~\ref{fig1}).
(a) We analyse the optimal dynamics for a Bosonic Josephson Junction (BJJ) \cite{lapert, gpe_qsl} with\add{in} the two-mode approximation\add{,} \add{where} one 
\add{mode (i.e., orbital) per well, solution of the}
\add{time-independent} Gross-Pitaevskii Equation (GPE), \add{is assumed}~\cite{bjj}. 
\add{(b) We investigate the mean-field scenario for which} the time-dependent GPE dynamics in the double well potential in \add{coordinate} space \add{is solved}, and finally (c) \add{we solve} 
the full many-body Schr\"odinger equation by means of the ab initio Multi-Configurational Time-Dependent Hartree method \add{for Bosons} (MCTDHB)\add{~\cite{mctdhb}}. 
\add{The latter approach allows us to \add{include}}  possible excitations to higher orbitals (\add{i.e.,} beyond \add{the} GPE)\add{, which are} responsible for quantum depletion and fragmentation.

In a linear two-level system the magnitude of the coupling $J$ between the two levels 
characterizes the QSL of a state-to-state transfer. 
\add{Hence, the} minimum time for the transfer between two orthogonal states is given by half the Rabi period\add{, that is,} $T_{QSL}^L=\pi \hbar/2J$~\cite{qsl_lit1,qsl_lit2}. 
\add{Such} time-optimal dynamics occurs along a geodesic on the Bloch sphere 
connecting the initial and the target state [see Fig.~\ref{fig1}(d)]. 
\add{Our main goal is to extend the geodesic interpretation of the QSL to interacting bosonic many-body quantum systems}.
\add{At the beginning} we  show that at the level of the two-mode BJJ model, which in the presence of  inter-particle interactions
becomes nonlinear~\cite{notetma}, the time-optimal dynamics \add{does follow} a geodesic too.
\add{This can be achieved by canceling the deviations from the geodesic path, due to the nonlinear interaction,} 
by \add{means of} analytically computable Compensating Control Pulses (CCP). 
We then study the range of validity of this strategy in  a double well \add{configuration} in \add{coordinate} space
where the two-mode
and GPE approximations
\add{may not hold anymore}. We show that the CCP 
performs very well even close to the QSL if \add{fragmentation, a genuine many-body effect, is neglected}. 
\add{We find, quite} surprisingly, \add{that} for strong interactions  
-- when the uncontrolled dynamics experience the self-trapping  -- the CCP leads to even faster speeds of \add{the} full transfer, 
taking efficiently advantage of the increase in the effective tunneling coupling. 
Finally \add{we show that}, when \add{large} deviations from the \add{ideal} transfer \add{efficiency} close to QSL \add{occur}, 
they \add{are} corrected by means of optimal control by
applying the Chopped RAndom Basis (CRAB) algorithm~\cite{crab}. 

\paragraph{QSL of a BJJ.}

The dynamics of the most general scenario of a quantum optimal control problem is described by the Hamiltonian $H(t)=H_d+ \sum_i c_i(t) H_i$, 
where $H_d$ represents the time-independentpart of the dynamics (drift), and $H_i$ are the control parts with control pulses $c_i(t)$. 
Here we focus on the task of optimizing the transfer from an initial state $\ket{\psi _0}$ to a target state $\ket{\psi _T}$. 
Optimal control looks for an optimal time dependence of the control pulses $c_i(t)$ such that a given figure of merit is minimized~\cite{qcbook}. 
For the latter we use the infidelity $\varepsilon=1-F^2$, where $F(\rho_T, \rho) = \operatorname{Tr} \left[\sqrt{\sqrt{\rho_T} \rho \sqrt{\rho_T}}\right]$ is the Uhlmann fidelity which measures 
the overlap of the two density matrices $\rho_T$ and $\rho$ \cite{fidelity-paper} and 
vanishes when the transfer to the target state is complete.
The typical questions that arise in optimal control are threefold: (i) is a complete transfer possible ($\varepsilon=0$)? 
(ii) \add{If} so, what is the minimum time (\add{i.e.,} the quantum speed limit) 
to achieve it? (iii) \add{What} are the optimal \add{control} pulses $c_i(t)$ that minimize the infidelity, the \add{transfer} time, or both? 
The answers to \add{these} questions depend on the specific quantum system~\cite{two-level,bjj_exp3, multi-level}. 
\add{Recently} formal expression\add{s} have been derived to compute the QSL \add{for a general Liouvillian evolution}~\cite{qslanalytics,qslanalytics2}.

The Hamiltonian of \add{a} linear two-level system reads: $H_L(t)=- \hbar J \sigma_x+ \hbar  D(t) \sigma_z$, where the drift term expresses the coupling strength $J$ between the two levels and $D(t)$ is 
the (controllable) detuning. 
A prototypical process that can be driven in such a system is a state-to-state transfer~\cite{qsl_exp1,two-level,bjj_exp3}. 
In Fig.~\ref{fig1} we represent the quantum state $|\psi(t)\rangle=e^{i\alpha(t)}\{\cos[\theta(t)/2]|\psi_0\rangle+\sin[\theta(t)/2]e^{i\phi(t)}|\psi_T\rangle\}$ on a Bloch sphere.
The speed of a quantum state moving in the Hilbert space is:  $\dot{s}(t)=2 \Delta  H(t)/\hbar$\add{,}
where $\Delta H(t)=\sqrt{\langle H^2(t) \rangle-\langle H(t)\rangle^2}$\add{,}
and $s$ is the geodesic distance between two states $|\psi \rangle$,$|\chi \rangle$,  defined as \add{$|\langle \chi|\psi \rangle|^2=\cos^2(s/2)$}~\cite{aharonov}. 
In a two-level system, the speed $\dot{s}(t)=\sqrt{\dot{\theta}(t)^2+\sin^2\theta(t)\dot{\phi}(t)^2}$  has two components: 
$\dot{\theta}(t)=J\sin\phi(t)$, which drives the state along the meridians of the Bloch sphere\add{,} 
and $\dot{\phi}(t)=D(t)+J\cot\theta(t)\cos\phi(t)$ along the parallels\add{.}
\add{In order to reach} the time-optimal transfer between north and south pole of the Bloch sphere \add{(see also Fig.~\ref{fig1})},  
only the speed term $\dot{\theta}(t)$ should be maximized: that is,  $\dot \theta = J$ at $\phi(t)=\pm \pi/2$, which implies $D(t)=0$.
Therefore the pulse given by the condition $D(t)=0$ \add{minimizes} the infidelity \add{(i.e.,} $\varepsilon\rightarrow 0$\add{)},  the \add{transfer} time $T_{QSL}^L$, 
and the path length  $S =\int\dot s \, dt =\pi$ following the geodesic ($\phi(t)$ is constant). 
In case constraints are present on the control pulse, e.g.
certain initial and final values of the detuning are assumed and can not be changed instantaneously, the optimal strategy \add{is still the one that} fulfils as much as possible the condition $\add{\vert} D\add{\vert}=0$, as shown numerically in \add{Ref.}~\cite{two-level}  and experimentally in Ref. ~\cite{qsl_exp1}.
We have assumed that $J$ is time-independent,  since, 
if $J$ can be varied, the time optimal solution is trivially \add{the one that} maximize\add{s} its value, \add{ and consequently} minimizes $T_{QSL}^L$. 

\add{Everything we have discussed so far applies to} time-dependent or time-independent, linear or nonlinear Hamiltonians, provided that the coupling 
term $J$ is $\phi$-independent. 
For instance\add{,} the BJJ Hamiltonian has an additional term with respect to the linear case $H_L$: $H_{BJJ}=H_L+\hbar[U+\Delta U \cos\theta(t)] \mathbb{I} +\hbar [\Delta U+U \cos\theta(t)]\sigma_z$,
where $U$ is the nonlinear interaction strength and $\Delta U$ the interaction detuning between the two modes (see \add{Supplemental Material} for details)~\cite{bjj}.
The nonlinear terms affect  the speed on the parallels so that in the corresponding equation of motion $D(t)$ has to be replaced by $\Delta(t)=D(t)+\Delta U + U \cos\theta(t)$.
Now even for $D(t)=0$ there is a non zero speed component that drives the state out of the geodesic path ($U \ne 0$ implies $\dot \phi \ne 0$), as shown by the trajectory depicted in Fig.~\ref{fig1} (red line). 

Now, in order to recover the time-optimal trajectory on a geodesic (green line) we impose the condition $\Delta(t)=0$ which here translates to~\cite{note0}:
 \begin{equation}
\label{CP}
D(t)=-\Delta U-U \cos\theta(t).
\end{equation}
Finding the solution of this equation appears not trivial since it is a nonlinear problem: $\cos\theta(t)$ depends on the parameters of the problem ($D(t)$\add{,} $U$, $\Delta U$) and the time evolution of $\phi(t)$ itself.
However, from the solution of the linear case, we know that if $\Delta(t)=0$ the system will move on a geodesic with maximum speed in a Rabi oscillation,  i.e. $\cos\theta(t)=\cos^2(Jt)$. 
In the following we \add{will} refer to this \add{pulse} as compensating control pulse \add{(CCP)} and investigate numerically its effectiveness and robustness. 

 Let us highlight that the CCP technique has various advantages for all experiments that work in the two-mode framework:
(a) It has a simple form that makes it applicable in experimental setups from Landau-Zenner transitions \cite{qsl_exp1} to double-well \cite{bjj_exp1,bjj_exp2,bjj_exp3} and internal  BJJ \cite{internalbjj} dynamics.
(b) It addresses the particularly difficult problem of such experiments, the motion on the meridians of the Bloch sphere, by overcoming the self-trapping mechanism.
Therefore it allows to reach any point on the Bloch sphere (also non-orthogonal to the initial one) at minimum time, since any phase-shift (motion on $\phi$) is easily achievable by setting an arbtrarily large detuning $D(t)$ or decoupling the modes.
To the extent that there is no constraint in $D(t)$ (an assumption valid also in \cite{qsl_exp1} and \cite{two-level}) the time needed for this phase shift can be made arbitrarily small.
(c) Our technique can nearly solve the problem of damping of the BJJ dynamics from which some experiments suffer \cite{bjj_exp2,interferometer} by setting the term which is responsible for it ($\dot{\phi}$) to zero \cite{damping}.
(d) It is finally a general concept on how to keep the phase constant and change only the imbalance of population,
which can be of a great help for reading, mapping and other features of interferometry processes \cite{interferometer} 

\paragraph{Numerical results.} 
We have verified numerically, within the two-mode approximation\add{,} that \add{the}
CCP  drives the system along a geodesic, efficiently canceling the nonlinear interactions\add{,} and \add{therefore achieving} the full transfer at $T_{QSL}$, as \add{it is also shown} 
in Fig.~\ref{fig1}  (green \add{path}). 
If there are constraints on the pulse, e.g. boundary conditions on the initial an final values of $D(t)$, then the condition of Eq.~\eqref{CP} cannot be fulfilled at every time, as required by the CCP scheme.
Nevertheless, one could then devise an
optimal control \add{pulse by means of numerical optimisation.} \add{A} typical result is reported in Fig.~\ref{fig1} (blue \add{path}) \add{which has been} 
obtained by \add{using the} CRAB \add{algorithm.} \add{Note that the dynamics generated by the constrained optimal control pulse approximately follows a trajectory similar to the one induced by the} CCP, \add{namely} moving at any time 
as close as possible to a geodesic. 

\begin{figure}
\includegraphics[height=6.5 cm, width=\columnwidth]{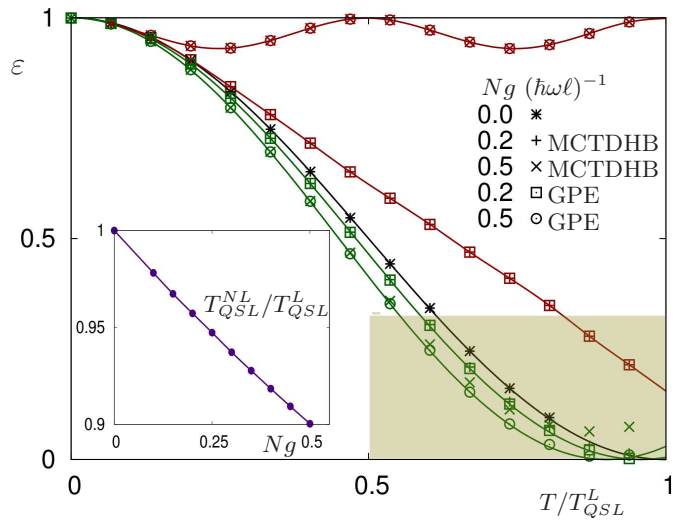} 
\put(-250,160){{\large $\varepsilon$} }
\put(-75,152){ $Ng$ $(\hbar \omega \ell)^{-1}$}
\put(-47,131){MCTDHB}
\put(-47,121){MCTDHB}
\put(-47,111){GPE}
\put(-47,101){GPE}
\put(-180,70){ $T_{QSL}^{NL}/T_{QSL}^{L}$}
\put(-158,17){ $Ng$}
\put(-50,-5){$T/T_{QSL}^L$}
\caption{Dynamics in a double-well: (a) Infidelity $\varepsilon$ as a function of the \add{transfer} time $T$ for representative values of the interaction strength $Ng$ 
(weak $Ng=0.2$  and strong $Ng=0.5$)
for uncontrolled dynamics (red \add{symbols} and curves) 
and for the CCP (green \add{symbols}) with fitting on a geodesic Rabi-oscillation behavior (green curves). 
\add{For solving the dynamics two methods have been used: GPE and MCTDHB.}
Black curve and asterisk symbols correspond to the non-interacting uncontrolled dynamics on the geodesic.
\add{The shadowed area in the lower righthand corner highlights the region close to $T_{QSL}$ (see also Fig.~\ref{fig3}).}
Inset: QSL time\add{,} $T_{QSL}^{NL}$\add{,} as a function of the nonlinear interaction strength $Ng$ 
compared to the non-interacting \add{(}linear\add{)} case $T_{QSL}^{L}$.\vspace{-0.6cm}
}
\label{fig2}
\end{figure}

We now go beyond the two-mode approximation \add{by} 
applying the previous findings to a \add{Bose-Einstein condensate (}BEC\add{)} in a one-dimensional double-well potential, 
the most common experimental realization of a\add{n atomic} BJJ\add{~\cite{bjj_exp1, bjj_exp2, bjj_exp3}}. 
In those setups it is possible to  
vary in time the tilt of the double well, which corresponds to \add{controlling} $D(t)$ in the two-mode model\add{. Our goal is} to obtain a \add{complete} 
transfer of the BEC\add{, initially prepared in the left well of the double well potential, to the right well}.
\add{In particular, we consider a} system composed by 
$N=100$ interacting bosons \add{trapped in a double well consisting of two separated harmonic traps of equal frequency $\omega$  and length  $\ell = \sqrt{\hbar/(m\omega)}$ \cite{kaspar}  (see Fig.~\ref{fig1} 
and the Supplemental Material technical details).}
\add{We have then applied the} different 
control schemes introduced before (CCP/CRAB), for various
interaction strengths $Ng$, \add{transfer} times $T$, and \add{approaches to solve the many-body quantum} dynamics (GPE/MCTDHB).

\add{To begin with, we} consider the GPE dynamics and in Fig.~\ref{fig2} we compare the infidelity for \add{a} symmetric (\add{unbiased}) double well potential (red curves)
\add{within the scenario where} the CCP on the tilt \add{has been applied} (green curves), 
as a function of the \add{transfer} time $T$ \add{as well as} for various interaction strengths $Ng$.
The time-independent scenario confirms the standard BJJ theory, which predicts a Rabi oscillation (on the geodesic) for vanishing \add{interaction strength} (black line), a full but delayed transfer 
for weak \add{interactions} $Ng=0.2 (\hbar \omega \ell)^{-1}$, and a nonlinear self-trapping \add{regime}, where less than half of the atoms are transferred to the right well, 
for strong interaction strengths  $Ng=0.5 (\hbar \omega \ell)^{-1}$ (see \add{also} Supplemental Material).
The \add{situation} changes drastically when we apply the CCP: 
for all cases the infidelity follows a $\cos^2(JT)$ behavior as all the fitting green lines in Fig.~\ref{fig2} show. 
These results demonstrate that the CCP drives the system along the geodesic, resulting in a successful time-optimal transfer within the GPE dynamics, for all values of $Ng$.
Moreover, the CCP not only compensates the nonlinearity recovering the linear single-particle dynamics (black line in Fig.~\ref{fig2}), but it increases the speed of the transfer
as demonstrated in the inset of Fig.~\ref{fig2}: a decrease of the $T_{QSL}$ for larger nonlinearities is reported there. 
This is \add{indeed} surprising\add{,} since it contrasts \add{with} the usual situation where repulsive interactions slow down the tunneling, 
and even completely prevent it in the case of strong nonlinear self-trapping \add{(}\add{see} red lines in Fig.~\ref{fig2}\add{)}. 
On the other hand, in a two-mode intuitive picture, the repulsion leads  to a broadening of the initial and target wave-functions, 
which increases their overlap, and thus their tunnel coupling $J$ (see sketches of Fig.~\ref{fig1} and Supplemental Material).
As a consequence, the speed enhancement of \add{the} tunneling transfer is obtained by the control pulses \add{that} cancel the delaying and self-trapping effects of the repulsion
and exploit its positive effects.

\begin{figure}
\includegraphics[height=4.5 cm, width=\columnwidth]{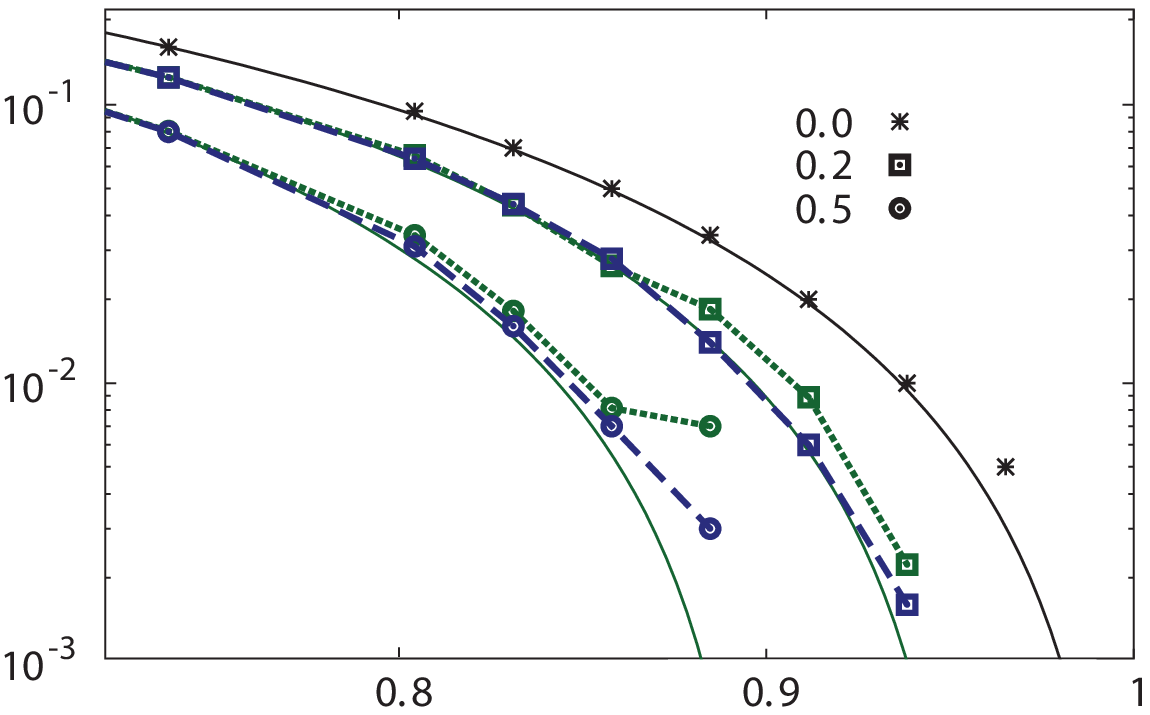}
\put(-20,118){(a) }
\put(-242,118){{\large $\varepsilon$} }
\put(-79,113){ $Ng$ $(\hbar \omega \ell)^{-1}$}
\\
\includegraphics[height=5.5 cm, width=\columnwidth]{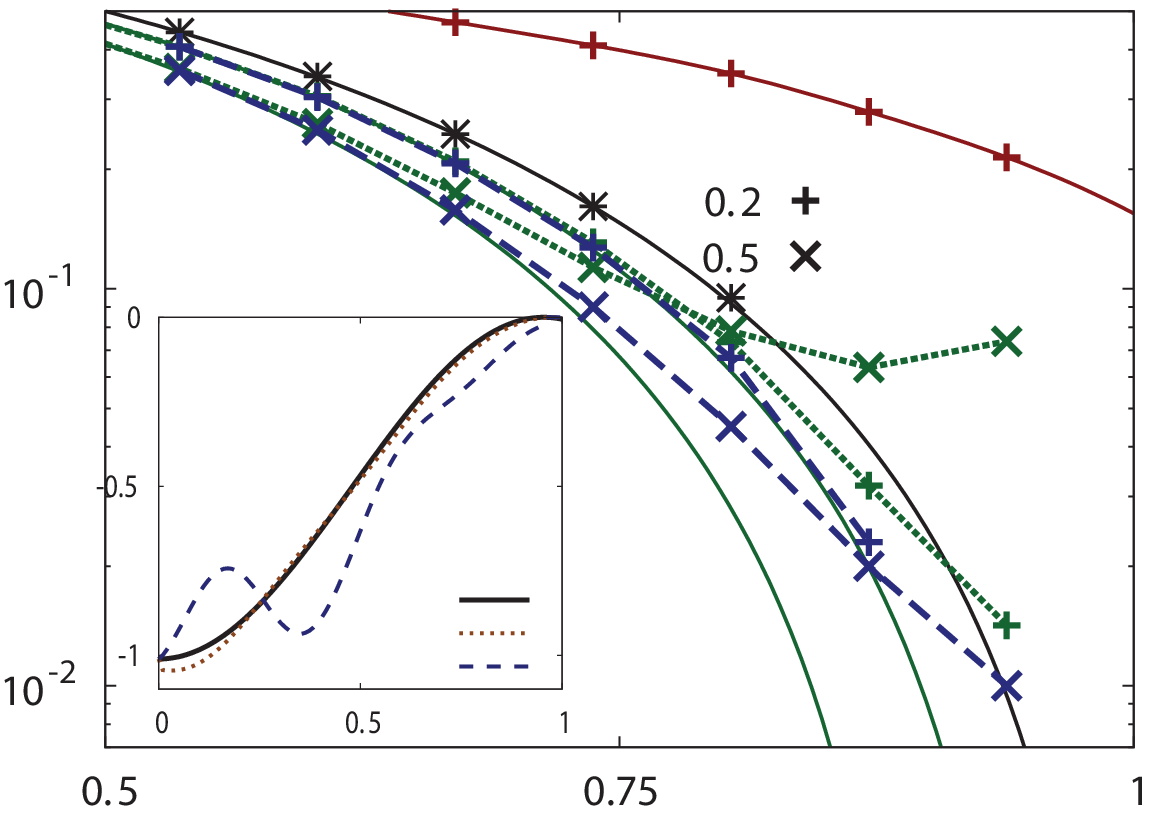} \
\put(-50,-5){$T/T_{QSL}^L$}
\put(-20,145){(b) }
\put(-242,148){{\large $\varepsilon$} }
\put(-101,125){ $Ng$ $(\hbar \omega \ell)^{-1}$}
\put(-190,100){ $D(t)/U$} 
\put(-172,41){CCP}
\put(-172,33){GPE}
\put(-195,26){MCTDHB}
\caption{Focus on the QSL region: Infidelities in log scale, for (a) GPE and (b) MCTDHB results (with same symbols as in Fig.~\ref{fig2}) of CCP pulses (green \add{symbols}), and CRAB optimizations (blue \add{symbols}).
Inset: Representative optimal control pulses for the tilt of the double-well near \add{the} QSL. \vspace{-0.6cm}
}
\label{fig3}
\end{figure}

Fig\add{ure}~\ref{fig3}(a) reports the same results, but \add{it} focuses on the challenging time regime close to $T_{QSL}$, where the deviations from the two-mode approximation \add{become significant}.
We note, however, that the CCPs still achieve quite \add{small values of the} infidelit\add{y} (green \add{symbols}), 
\add{although they} do not reach the expected values \add{corresponding to} a perfect geodesic trajectory (green curves).
We then resort to CRAB optimizations, which correct these discrepancies and produce optimal \add{control} pulses that result in values of $\varepsilon$ much closer to the curve (blue \add{symbols}). 
We \add{thus} conclude that the geodesic behavior obtained by \add{the} CCP (with possibly an additional optimization) 
enables time-optimal GPE dynamics in a double-well.

\add{Up to now, we have investigated the dynamics of the bosons in the framework of mean field GPE theory.
For small particle numbers such description might become inadequate, 
and in the limit of large repulsion between the bosons  fails completely. 
Thus we investigate the driven dynamics of the bosons also by means of the MCTDHB method, 
which enables us to solve the many-body Schr\"odinger equation exactly, up to a target numerical accuracy that can be set a priori.} 
In \add{Ref.}~\cite{kaspar} it has been shown in a similar time-independent BJJ double-well potential, \add{that}
deviations from the GPE predictions for all time scales, and especially for strong interactions, might occur. 
Indeed, deviations induced by many-body fragmentation and depletion 
depend on several factors: number of atoms, interaction strength, 
dimensionality, total time \add{evolution}, strength of the control driving, potential profile.
In Fig.~\ref{fig2} one can clearly see that the undriven dynamics calculated \add{with} MCTDHB (red cross and plus symbols) do not have significant deviations from the red curves obtained \add{via} GPE.
However, when we apply the CCP,  MCTDHB
(green cross and plus symbols) and \add{the} GPE dynamics agree only for short time scales,
while close to the QSL [see also Fig.~\ref{fig3}(b)] 
and for strong interaction strengths there is \add{a} substantial \add{disagreement} (of up to one order of magnitude).
This discrepancy occurs as for long times, strong driving, and strong couplings the fragmentation becomes not negligible (depletion of the condensate orbital \add{of} the order of $10\%$).
\add{Hence, for these particular situations} we need to rely on optimal control \add{in order} to \add{further improve the transfer efficiency}.
To this aim\add{,} we applied \add{the} CRAB optimization to \add{the} MCTDHB simulation \add{for transfer times} in the vicinity of the QSL. 
We succeeded to reduce the infidelities from $\varepsilon\approx 10 \%$ to  $\varepsilon\approx 1 \%$ (see Fig.~\ref{fig3}(b) blue points).
In the inset of Fig.~\ref{fig3}(b) we present representative optimal pulses for the tilt of the double well close to \add{the} QSL\add{. As it is shown,} there is \add{a large} difference between CRAB optimizations 
performed with the GPE or exact many-body MCTDHB simulations.

In conclusion we have shown that the repulsive interaction \add{between the particles} \add{is} responsible \add{for} slowing down the dynamics of a bosonic Josephson junction, and also precludes a \add{highly efficient} state transfer. 
We have provided the conditions to design a compensating \add{control} pulse that \add{eliminates} the detrimental effects of the interaction and drive the system on a geodesic path at the quantum speed limit.
Our driving scheme allows for an increase of the transfer speed for stronger interaction strengths as it exploits the constructive effects of the interaction. 
Close to the QSL \add{and for large interaction strengths}, \add{the} mean field description does not accurately account for the dynamics, and a simple compensating \add{control} pulse strategy \add{is not sufficient to achieve high transfer efficiency.} \add{Given this,} we have to resort to 
optimal control of the \add{many-body Schr\"odinger equation}, \add{which can be efficiently solved} by \add{the} ab-initio MCTDHB \add{method}.
Our study introduced the concepts to characterize the QSL for many-body systems at all levels of complexity of the description of the 
problem (two-mode, nonlinear mean-field, full many-body), and provided optimal control schemes to achieve the state-to-state transfer at minimum time. 
 Our control pulses are simple, feasible and easily applicable to a variety of linear nonlinear and many-body systems in the laboratory, and it can provide a time-optimal solution to common issues like self-trapping and damping of oscillations, which limit the experimental possibilities of atom transfer and interferometry.

\paragraph{Acknowledgements\add{.}} We acknowledge the support of the EU funded FET project QUIBEC, SIQS, QUAINT (GA $\mathrm{N^o}$  297861), the German Research Foundation (DFG) via the SFB/TRR21,
\add{and} the excellence cluster Hamburg CUI of the Forschungsgemeinschaft, the computational time within BwGrid and HLRS computational facilities and DFG grant number CE-10/52-1.  We thank the group of J\"org Schmiedmayer for inspiring discussions.


\begin{thebibliography}{10}
\bibitem{note0} 
For $U,\Delta U=0$ we recover the pulse for the linear system $D(t)=0$. Also we have assumed here control only on $D(t)$ but if also the other parameters are 
time-dependent $U(t),\Delta U(t)$, still the condition for the optimal protocol remains the same.
\end{thebibliography}
\end{document}